\PassOptionsToPackage{unicode}{hyperref}
\PassOptionsToPackage{hyphens}{url}
\PassOptionsToPackage{dvipsnames,svgnames,x11names}{xcolor}
\documentclass{article}
\usepackage{xcolor}
\usepackage{amsmath,amssymb}
\usepackage{iftex}
  \usepackage[T1]{fontenc}
  \usepackage[utf8]{inputenc}
  \usepackage{textcomp} 
\usepackage{lmodern}
\IfFileExists{upquote.sty}{\usepackage{upquote}}{}
\IfFileExists{microtype.sty}{
  \usepackage[]{microtype}
  \UseMicrotypeSet[protrusion]{basicmath} 
}{}
\makeatletter
\@ifundefined{KOMAClassName}{
  \IfFileExists{parskip.sty}{%
    \usepackage{parskip}
  }{
    \setlength{\parindent}{0pt}
    \setlength{\parskip}{6pt plus 2pt minus 1pt}}
}{
  \KOMAoptions{parskip=half}}
\makeatother
\makeatletter
\ifx\paragraph\undefined\else
  \let\oldparagraph\paragraph
  \renewcommand{\paragraph}{
    \@ifstar
      \xxxParagraphStar
      \xxxParagraphNoStar
  }
  \newcommand{\xxxParagraphStar}[1]{\oldparagraph*{#1}\mbox{}}
  \newcommand{\xxxParagraphNoStar}[1]{\oldparagraph{#1}\mbox{}}
\fi
\ifx\subparagraph\undefined\else
  \let\oldsubparagraph\subparagraph
  \renewcommand{\subparagraph}{
    \@ifstar
      \xxxSubParagraphStar
      \xxxSubParagraphNoStar
  }
  \newcommand{\xxxSubParagraphStar}[1]{\oldsubparagraph*{#1}\mbox{}}
  \newcommand{\xxxSubParagraphNoStar}[1]{\oldsubparagraph{#1}\mbox{}}
\fi
\makeatother

\usepackage{longtable,booktabs,array}
\usepackage{calc} 
\usepackage{etoolbox}
\makeatletter
\patchcmd\longtable{\par}{\if@noskipsec\mbox{}\fi\par}{}{}
\makeatother
\IfFileExists{footnotehyper.sty}{\usepackage{footnotehyper}}{\usepackage{footnote}}
\makesavenoteenv{longtable}
\usepackage{graphicx}
\makeatletter
\newsavebox\pandoc@box
\newcommand*\pandocbounded[1]{
  \sbox\pandoc@box{#1}%
  \Gscale@div\@tempa{\textheight}{\dimexpr\ht\pandoc@box+\dp\pandoc@box\relax}%
  \Gscale@div\@tempb{\linewidth}{\wd\pandoc@box}%
  \ifdim\@tempb\p@<\@tempa\p@\let\@tempa\@tempb\fi
  \ifdim\@tempa\p@<\p@\scalebox{\@tempa}{\usebox\pandoc@box}%
  \else\usebox{\pandoc@box}%
  \fi%
}
\def\fps@figure{htbp}
\makeatother

\NewDocumentCommand\citeproctext{}{}

\makeatletter
 \let\@cite@ofmt\@firstofone
 \def\@biblabel#1{}
 \def\@cite#1#2{{#1\if@tempswa , #2\fi}}
\makeatother
\newlength{\cslhangindent}
\newlength{\csllabelwidth}
\newenvironment{CSLReferences}[2] 
 {\begin{list}{}{%
  \setlength{\itemindent}{0pt}
  \setlength{\leftmargin}{0pt}
  \setlength{\parsep}{0pt}
  \ifodd #1
   \setlength{\leftmargin}{\cslhangindent}
   \setlength{\itemindent}{-1\cslhangindent}
  \fi
  \setlength{\itemsep}{#2\baselineskip}}}
 {\end{list}}
\usepackage{calc}

\providecommand{\tightlist}{%
  \setlength{\itemsep}{0pt}\setlength{\parskip}{0pt}}

\usepackage{arxiv}
\usepackage{orcidlink}
\usepackage{amsmath}
\usepackage[T1]{fontenc}
\makeatletter
\@ifpackageloaded{caption}{}{\usepackage{caption}}
\AtBeginDocument{%
\ifdefined\contentsname
  \renewcommand*\contentsname{Table of contents}
\else
  \newcommand\contentsname{Table of contents}
\fi
\ifdefined\listfigurename
  \renewcommand*\listfigurename{List of Figures}
\else
  \newcommand\listfigurename{List of Figures}
\fi
\ifdefined\listtablename
  \renewcommand*\listtablename{List of Tables}
\else
  \newcommand\listtablename{List of Tables}
\fi
\ifdefined\figurename
  \renewcommand*\figurename{Figure}
\else
  \newcommand\figurename{Figure}
\fi
\ifdefined\tablename
  \renewcommand*\tablename{Table}
\else
  \newcommand\tablename{Table}
\fi
}
\@ifpackageloaded{float}{}{\usepackage{float}}
\floatstyle{ruled}
\@ifundefined{c@chapter}{\newfloat{codelisting}{h}{lop}}{\newfloat{codelisting}{h}{lop}[chapter]}
\floatname{codelisting}{Listing}

\makeatother
\makeatletter
\@ifpackageloaded{caption}{}{\usepackage{caption}}
\@ifpackageloaded{subcaption}{}{\usepackage{subcaption}}
\makeatother
\usepackage{bookmark}
\IfFileExists{xurl.sty}{\usepackage{xurl}}{} 
\hypersetup{
  pdftitle={TrustFlow: Topic-Aware Vector Reputation Propagation for Multi-Agent Ecosystems},
  pdfauthor={Volodymyr Seliuchenko},
  pdfkeywords={reputation, trust propagation, multi-agent
systems, PageRank, topic-sensitive, embedding space},
  colorlinks=true,
  linkcolor={blue},
  filecolor={Maroon},
  citecolor={Blue},
  urlcolor={Blue},
  pdfcreator={LaTeX via pandoc}}

\renewcommand{\today}{2026-03-01}
\newcommand{\runninghead}{A Preprint }
\title{TrustFlow: Topic-Aware Vector Reputation Propagation for
Multi-Agent Ecosystems}
\author{\textbf{Volodymyr Seliuchenko}\\\\robutler.ai\\\\}
\date{2026-03-01}
\begin{document}
\maketitle
\begin{abstract}
We introduce TrustFlow, a reputation propagation algorithm that assigns
each software agent a multi-dimensional reputation vector rather than a
scalar score. Reputation is propagated through an interaction graph via
topic-gated transfer operators that modulate each edge by its content
embedding, with convergence to a unique fixed point guaranteed by the
contraction mapping theorem. We develop a family of Lipschitz-1 transfer
operators and composable information-theoretic gates that achieve up to
98\% multi-label Precision@5 on dense graphs and 78\% on sparse ones. On
a benchmark of 50 agents across 8 domains, TrustFlow resists sybil
attacks, reputation laundering, and vote rings with at most 4
percentage-point precision impact. Unlike PageRank and Topic-Sensitive
PageRank, TrustFlow produces vector reputation that is directly
queryable by dot product in the same embedding space as user queries.
\end{abstract}
{\bfseries \emph Keywords}
\def\sep{\textbullet\ }
reputation \sep trust propagation \sep multi-agent
systems \sep PageRank \sep topic-sensitive \sep 
embedding space

\section{Introduction}\label{introduction}

The emergence of autonomous software agents that interact, delegate
tasks, and transact on behalf of users creates a fundamental trust
problem: how does a user---or another agent---select the right agent for
a task from a marketplace of millions? Traditional approaches (star
ratings, manual curation, call counts) fail at scale because they are
easily gamed, domain-agnostic, and cannot propagate transitive trust.

PageRank (Brin and Page 1998) demonstrated that the link structure of
the web encodes a useful notion of importance: a page is important if
important pages link to it. Topic-Sensitive PageRank (TSPR) (Haveliwala
2002) extended this insight by computing multiple PageRank scores, each
biased toward a different topic, to yield query-sensitive importance
scores. Both algorithms operate on a single, topic-independent
transition matrix \(M\) and produce scalar rankings.

Agent reputation, however, has structural properties absent from
web-page ranking that demand a fundamentally different formulation:

\begin{enumerate}
\def\labelenumi{\arabic{enumi}.}
\tightlist
\item
  \textbf{Multi-domain expertise.} An agent may be expert in both
  medicine and data science. A scalar reputation score cannot represent
  this; a vector can.
\item
  \textbf{Content--competence decoupling.} Web pages are inherently
  content-bearing---text and links \emph{are} the signal. Agents may be
  pure service providers; trust must derive from interaction history,
  not static content.
\item
  \textbf{Economic signals.} When agent \(A\) delegates a paid task to
  agent \(B\), the economic commitment is a trust signal absent from
  hyperlink graphs.
\item
  \textbf{Adversarial moderation.} Malicious agents may attempt sybil
  attacks, reputation laundering, or vote rings. The system must resist
  these while maintaining fairness.
\item
  \textbf{Directional discovery.} Users search for agents by
  natural-language queries. Reputation should be queryable in the same
  vector space as queries, enabling a single dot-product retrieval.
\end{enumerate}

TrustFlow addresses all five through a unified framework of vector
reputation propagation with topic-gated trust transfer. Our
contributions:

\begin{enumerate}
\def\labelenumi{\arabic{enumi}.}
\tightlist
\item
  \textbf{Topic-aware vector reputation propagation} in both discrete
  (\(N \times D\) matrix) and continuous (\(N \times E\) embedding)
  formulations, with convergence guarantees via contraction mapping
  (\S3).
\item
  \textbf{A family of Lipschitz-1 transfer operators}---projection,
  squared gating, scalar-gated, and hybrid---that modulate reputation
  flow by interaction content, with composable information-theoretic
  gates (KL-divergence, entropy, confidence) and empirically
  characterized precision--information tradeoffs (\S3.3).
\item
  \textbf{Negative trust edges} for moderation flags with joint
  convergence guarantees (\S3.6).
\item
  \textbf{Comprehensive evaluation} demonstrating structural resilience
  to four attack classes, with \({\leq}4\)pp P@5 impact across all
  transfer operators (\S5).
\end{enumerate}

\begin{center}\rule{0.5\linewidth}{0.5pt}\end{center}

\section{Preliminaries}\label{preliminaries}

We briefly review the mathematical foundations on which TrustFlow
builds.

\subsection{PageRank}\label{pagerank}

Let \(N\) be the number of pages on the web and let \(M\) be the
\(N \times N\) row-stochastic transition matrix of the web graph, where
\(M_{ij} = 1/\text{outdeg}(i)\) if page \(i\) links to page \(j\). The
PageRank vector \(\mathbf{r}\) is the stationary distribution of the
random walk:

\begin{equation}\phantomsection\label{eq-pagerank}{\mathbf{r} = \alpha\, M^T \mathbf{r} + (1 - \alpha)\, \mathbf{v}}\end{equation}

where \(\alpha \in (0,1)\) is the damping factor and \(\mathbf{v}\) is
the teleportation vector (uniform in standard PageRank). The
teleportation term ensures ergodicity---stochasticity, irreducibility,
and aperiodicity---guaranteeing convergence to a unique fixed point
(Langville and Meyer 2006). Each entry \(r_i\) is a scalar importance
score.

\subsection{Topic-Sensitive PageRank}\label{topic-sensitive-pagerank}

Haveliwala (2002) observed that a single PageRank vector cannot capture
topic-dependent importance. TSPR computes \(K\) biased PageRank vectors
\(\{\mathbf{r}^{(k)}\}_{k=1}^K\), each using a topic-specific
teleportation vector \(\mathbf{v}^{(k)}\) that concentrates probability
mass on pages known to belong to topic \(k\):

\begin{equation}\phantomsection\label{eq-tspr}{\mathbf{r}^{(k)} = \alpha\, M^T \mathbf{r}^{(k)} + (1 - \alpha)\, \mathbf{v}^{(k)}}\end{equation}

At query time, the query is classified into topics and the final score
is a linear combination \(\sum_k p(k \mid q)\, r_i^{(k)}\). Crucially,
the transition matrix \(M\) is the \emph{same} for all topics---TSPR
cannot distinguish ``A links to B in a medical context'' from ``A links
to B in a coding context.''

\subsection{Notation}\label{notation}

\begin{longtable}[]{@{}
  >{\raggedright\arraybackslash}p{(\linewidth - 2\tabcolsep) * \real{0.4211}}
  >{\raggedright\arraybackslash}p{(\linewidth - 2\tabcolsep) * \real{0.5789}}@{}}
\caption{Notation summary.}\label{tbl-notation}\tabularnewline
\toprule\noalign{}
\begin{minipage}[b]{\linewidth}\raggedright
Symbol
\end{minipage} & \begin{minipage}[b]{\linewidth}\raggedright
Definition
\end{minipage} \\
\midrule\noalign{}
\endfirsthead
\toprule\noalign{}
\begin{minipage}[b]{\linewidth}\raggedright
Symbol
\end{minipage} & \begin{minipage}[b]{\linewidth}\raggedright
Definition
\end{minipage} \\
\midrule\noalign{}
\endhead
\bottomrule\noalign{}
\endlastfoot
\(N\) & Number of agents \\
\(E\) & Embedding dimensionality (e.g., 384 for E5) \\
\(D\) & Number of discrete domains \\
\(R[i] \in \mathbb{R}^E\) & Reputation vector of agent \(i\) \\
\(e_{ij} \in \mathbb{R}^E\) & Unit interaction embedding for edge
\(i \to j\) \\
\(T[j] \in \mathbb{R}^E\) & Teleportation prior of agent \(j\) \\
\(C[j] \in \mathbb{R}^E\) & Exogenous authority injection for agent
\(j\) \\
\(\alpha\) & Damping factor (default 0.85) \\
\(w(i \to j)\) & Row-normalized edge weight \\
\(M_d\) & Domain-conditioned transition matrix for domain \(d\) \\
\(\odot\) & Element-wise (Hadamard) product \\
\end{longtable}

\begin{center}\rule{0.5\linewidth}{0.5pt}\end{center}

\section{The TrustFlow Algorithm}\label{the-trustflow-algorithm}

\subsection{Overview}\label{overview}

TrustFlow generalizes PageRank along two axes:

\begin{itemize}
\tightlist
\item
  \textbf{Scalar \(\to\) vector.} Each agent's reputation is a vector
  \(R[i] \in \mathbb{R}^E\), not a scalar. The direction encodes the
  agent's expertise profile; the magnitude encodes accumulated trust.
\item
  \textbf{Single transition matrix \(\to\) topic-gated transfer.} Rather
  than TSPR's approach of biasing the teleportation vector while sharing
  a single \(M\), TrustFlow gates each edge's reputation transfer by the
  interaction's \emph{content embedding}, producing topic-dependent
  transfer without requiring pre-defined topic categories.
\end{itemize}

We present the continuous formulation first, then the discrete
specialization.

\subsection{Continuous Formulation}\label{continuous-formulation}

Each agent \(i\) has a reputation vector \(R[i] \in \mathbb{R}^E\) that
lives in the same embedding space as interaction content and discovery
queries. Each directed edge \((i \to j)\) carries a unit interaction
embedding \(e_{ij} \in \mathbb{R}^E\) derived from the content of their
interaction. The core iteration is:

\begin{equation}\phantomsection\label{eq-trustflow}{R_{new}[j] = \alpha \sum_{i} w(i \!\to\! j)\; f(R[i],\, e_{ij}) \;+\; (1-\alpha)\, T[j] \;+\; C[j]}\end{equation}

where:

\begin{itemize}
\tightlist
\item
  \(f(R[i], e_{ij})\) is a \textbf{topic-gated transfer operator} that
  modulates the sender's reputation by the interaction embedding. We
  study a family of such operators (\S3.3)---projection, element-wise
  squared gating, Hadamard relu, scalar-gated, and hybrid---each
  satisfying a Lipschitz-1 bound that guarantees convergence.
\item
  \(w(i \to j)\) is the row-normalized edge weight. Raw weights
  incorporate interaction frequency and optional payment delegation
  multipliers (e.g., \(\mu = 3\) for economically backed edges); row
  normalization is applied afterward so that \(\sum_j w(i \to j) = 1\)
  for each sender \(i\), preserving the contraction guarantee regardless
  of multiplier magnitude.
\item
  \(T[j] \in \mathbb{R}^E\) is the \textbf{teleportation prior}, playing
  the same structural role as \(\mathbf{v}\) in
  Equation~\ref{eq-pagerank}: scaled by \((1-\alpha)\), it ensures
  ergodicity (stochasticity, irreducibility, aperiodicity) exactly as in
  PageRank's Google Matrix. \(T[j]\) is derived from agent \(j\)'s
  public content embeddings, weighted by engagement and quality scores.
\item
  \(C[j] \in \mathbb{R}^E\) is the \textbf{exogenous authority
  injection}, an additive term that absorbs external reputation signals
  (content engagement, web authority, economic intent, cross-platform
  reputation) into the embedding space. Because \(C\) is not scaled by
  \((1-\alpha)\), the operator can tune exogenous signal strength
  independently of the damping factor; an alternative formulation
  couples \(C\) under \((1-\alpha)\) alongside \(T\) when exogenous
  signals should naturally diminish as graph evidence accumulates.
\end{itemize}

The iteration begins with \(R_0[j] = T[j] + C[j]\) and converges to a
unique fixed point by the contraction mapping theorem (\S3.5).

\textbf{Blind edges.} Edges whose content is not available (e.g.,
encrypted API calls) are called \emph{blind edges}. Because no
interaction text can be embedded, blind edges carry less signal than
labeled ones and are discounted (weighted at \(0.3\times\) in our
experiments). Their quality can be improved with proxy embeddings: in
this work we use \(e_{ij} = \text{avg}(p_i, p_j)\), the mean of the two
agents' profile embeddings, which assumes the interaction concerns a
topic between their areas of expertise. A learned model conditioned on
richer caller--callee features (capability overlap, historical call
patterns, task metadata) could predict a more accurate proxy and is a
direction for future work. We evaluate the averaging proxy in \S5.

\textbf{Unnormalized reputation.} We retain full magnitude during
iteration: \(R[i]\) is not L2-normalized between steps. An agent that
accumulates reputation from many high-quality interactions in a given
direction will have a larger component in that direction. The L2 norm
\(\|R[i]\|\) naturally reflects total accumulated reputation---analogous
to how PageRank propagates more authority from high-scoring
nodes---while the direction \(R[i]/\|R[i]\|\) encodes the expertise
profile. An alternative, per-iteration normalization variant constrains
all vectors to the unit sphere; we compare both in \S5.3.

\textbf{Comparison with PageRank and TSPR.} Equation~\ref{eq-trustflow}
generalizes Equation~\ref{eq-pagerank} in three ways: (i) scalar \(r_i\)
becomes vector \(R[i]\); (ii) the uniform transfer \(M^T \mathbf{r}\)
becomes the topic-gated transfer \(\sum_i w \cdot f(R[i], e_{ij})\); and
(iii) the teleportation \(\mathbf{v}\) is supplemented by an independent
exogenous injection \(C\). Unlike TSPR (Equation~\ref{eq-tspr}), which
biases only the teleportation vector while retaining a single
topic-independent \(M\), TrustFlow makes every edge's transfer depend on
the interaction's semantic content.

\subsection{Transfer Operator Family}\label{sec-operators}

The transfer operator \(f\) in Equation~\ref{eq-trustflow} determines
how reputation is modulated by the interaction embedding before
propagation. We study five base operators, each satisfying a Lipschitz-1
bound (ensuring convergence, \S3.5) but occupying a different point in
the precision--information tradeoff space.

\textbf{Projection}: \(f(R, e) = \sigma(R \cdot e) \cdot e\), where
\(\sigma = \max(0, \cdot)\). The output is always parallel to \(e_{ij}\)
(rank 1), confining transfer strictly to the interaction topic
direction. This provides maximum cross-domain isolation but collapses
directional information on content-free (blind) edges
(\(\cos \approx 0.004\) for uniform \(e\)).

\textbf{Squared gating}: \(f(R, e) = R \odot e^2\). Each dimension of
the sender's reputation is gated by the squared activation of the
interaction embedding in that dimension. Because \(e_k^2 \geq 0\), the
gating is always non-negative, acting as a spectral filter. On blind
edges (\(e = \frac{1}{\sqrt{E}}\mathbf{1}\)), the output is
\(R/E\)---the sender's full profile with perfect directional
preservation (\(\cos(\text{output}, R) = 1.0\)).

\textbf{Scalar-gated}: \(f(R, e) = \sigma(\hat{R} \cdot e) \cdot R\),
where \(\hat{R} = R/\|R\|\) and \(\sigma(x) = \min(\max(0, x), 1)\) is a
clamped relu. The clamp ensures \(\sigma \in [0,1]\), so
\(\|f(R_1, e) - f(R_2, e)\| \leq \|R_1 - R_2\|\), satisfying the
Lipschitz-1 bound. A single go/no-go gate based on overall cosine
alignment, transferring the full profile when the gate opens. Excels on
dense labeled graphs but degrades on blind-heavy graphs where the gate
closes.

\textbf{Hadamard relu}: \(f(R, e) = \max(0, R \odot e)\). A variant of
squared gating that applies element-wise relu instead of squaring.
Unlike squared gating, relu clips the \({\sim}52\%\) of dimensions where
dense E5 embeddings have negative components, losing information.

\textbf{Hybrid}: Per-edge operator selection based on content
availability (e.g., squared gating for content-free edges, projection
for content-rich edges), or a convex interpolation
\(\gamma \cdot f_\text{proj} + (1-\gamma) \cdot f_\text{sq}\).

\begin{longtable}[]{@{}
  >{\raggedright\arraybackslash}p{(\linewidth - 8\tabcolsep) * \real{0.1266}}
  >{\centering\arraybackslash}p{(\linewidth - 8\tabcolsep) * \real{0.1646}}
  >{\centering\arraybackslash}p{(\linewidth - 8\tabcolsep) * \real{0.2532}}
  >{\centering\arraybackslash}p{(\linewidth - 8\tabcolsep) * \real{0.2911}}
  >{\centering\arraybackslash}p{(\linewidth - 8\tabcolsep) * \real{0.1646}}@{}}
\caption{Transfer operator
properties.}\label{tbl-operators}\tabularnewline
\toprule\noalign{}
\begin{minipage}[b]{\linewidth}\raggedright
Operator
\end{minipage} & \begin{minipage}[b]{\linewidth}\centering
Output rank
\end{minipage} & \begin{minipage}[b]{\linewidth}\centering
Blind preservation
\end{minipage} & \begin{minipage}[b]{\linewidth}\centering
Cross-domain isolation
\end{minipage} & \begin{minipage}[b]{\linewidth}\centering
Convergence
\end{minipage} \\
\midrule\noalign{}
\endfirsthead
\toprule\noalign{}
\begin{minipage}[b]{\linewidth}\raggedright
Operator
\end{minipage} & \begin{minipage}[b]{\linewidth}\centering
Output rank
\end{minipage} & \begin{minipage}[b]{\linewidth}\centering
Blind preservation
\end{minipage} & \begin{minipage}[b]{\linewidth}\centering
Cross-domain isolation
\end{minipage} & \begin{minipage}[b]{\linewidth}\centering
Convergence
\end{minipage} \\
\midrule\noalign{}
\endhead
\bottomrule\noalign{}
\endlastfoot
Squared gating & Up to \(E\) & \(\cos = 1.0\) & Soft (per-dim filter) &
Always \\
Projection & 1 & \(\cos \approx 0\) & Maximum & Always \\
Hadamard relu & Up to \(E\) & \(\cos = 0.71\) & Moderate (relu clips) &
Always \\
Scalar-gated & Up to \(E\) & Full (when gate opens) & Binary & Always
(\(\sigma\) clamped) \\
Hybrid & Up to \(E\) & Tunable & Tunable & Always \\
\end{longtable}

All operators are Lipschitz-1. For squared gating:
\(\|(R_1 - R_2) \odot e^2\| \leq \|R_1 - R_2\| \cdot \|e^2\|_\infty \leq \|R_1 - R_2\|\)
since \(\|e\|_2 = 1\) implies \(|e_k| \leq 1\). For projection:
projection onto a unit vector is nonexpansive. Combined with
\(\alpha < 1\), all variants converge.

\textbf{KL-divergence gating.} Any transfer operator can be composed
with an information-theoretic gate that suppresses cross-domain
reputation leakage. The gated transfer replaces \(f(R[i], e_{ij})\) with
\(\text{gate}(i, e_{ij}) \cdot f(R[i], e_{ij})\), where:

\begin{equation}\phantomsection\label{eq-kl}{\text{gate}(i, e_{ij}) = \exp\bigl(-\lambda \cdot D_{KL}(p_\text{int} \| p_\text{rep}[i])\bigr)}\end{equation}

Here \(p_\text{int}\) is the interaction's topic distribution and
\(p_\text{rep}[i]\) is agent \(i\)'s reputation treated as a
distribution. When the interaction aligns with the sender's expertise
(\(D_{KL} \approx 0\)), the gate \(\approx 1\); when off-topic, the gate
decays exponentially. In the continuous formulation, an efficient cosine
proxy avoids the softmax conversion:
\(\text{gate}(i, e_{ij}) = \exp(-\lambda \cdot \sin^2\theta)\), where
\(\theta\) is the angle between \(R[i]\) and \(e_{ij}\). The gate is
bounded in \([0,1]\) and Lipschitz-continuous in \(R\), preserving the
contraction guarantee, though the composite contraction constant
increases slightly, requiring more iterations to converge.

\textbf{Additional gating mechanisms.} The gating architecture is
modular: KL-divergence is one instance of a family of composable gates,
all bounded in \([0,1]\) and convergence-preserving. Other members
include: (i) an \emph{entropy gate} \(\exp(-\mu \cdot H(p_\text{int}))\)
that suppresses unfocused or spam-like interactions with high topic
entropy; (ii) a \emph{magnitude-ratio gate} that measures the fraction
of the sender's reputation concentrated in the interaction direction,
penalizing incidental transfers far from the sender's core profile; and
(iii) a \emph{confidence gate} that discounts interactions where the
embedding model reports low certainty (particularly useful for blind
edges). Gates compose
multiplicatively---\(\text{gate} = \text{gate}_\text{KL} \cdot \text{gate}_\text{entropy} \cdot \text{gate}_\text{conf}\)---providing
layered defense against different forms of cross-domain leakage. We
evaluate KL-divergence gating in \S5.4; the remaining gates are left to
future work.

\subsection{Discrete Specialization}\label{discrete-specialization}

When \(D\) pre-defined domains are available (e.g., medicine, law,
finance), each interaction is classified into top-\(k\) domains via soft
assignment against domain centroids. This produces \(D\) separate
transition matrices \(M_d\), each row-stochastic, capturing the fraction
of interactions between agents relevant to domain \(d\). The per-domain
iteration is:

\begin{equation}\phantomsection\label{eq-discrete}{R_{new}[:,d] = \alpha\, M_d^T\, R[:,d] + (1-\alpha)\, T[:,d] + C[:,d]}\end{equation}

This is the key structural difference from TSPR: where TSPR uses the
\emph{same} \(M\) for all topics and varies only the teleportation
vector, TrustFlow uses \emph{different} \(M_d\) per domain. When agent
\(A\) calls agent \(B\) for a medical task, this edge appears with high
weight in \(M_\text{med}\) and low weight in \(M_\text{code}\). TSPR
cannot make this distinction.

Each per-domain iteration is an independent PageRank with
domain-specific structure, converging in
\(O(\log(1/\varepsilon)/\log(1/\alpha))\) iterations (approximately 11
for \(\alpha = 0.85\), \(\varepsilon = 10^{-4}\)) (Langville and Meyer
2006).

\textbf{Continuous vs.~discrete.} The discrete formulation requires a
pre-defined domain taxonomy and cannot represent cross-domain expertise
without dilution. A biostatistics agent must split its reputation mass
across medicine and data-science bins. The continuous formulation
eliminates this bottleneck: a 384-dimensional vector can simultaneously
align with both domain centroids. The discrete formulation is a
specialization of the continuous one---obtained by replacing the
embedding space \(\mathbb{R}^E\) with a domain indicator space
\(\mathbb{R}^D\) and the content-gated transfer with domain-conditioned
matrices.

\subsection{Convergence}\label{sec-convergence}

\textbf{Theorem 1} (Convergence). \emph{For any Lipschitz-1 transfer
operator \(f\) and damping factor \(\alpha \in (0,1)\), the TrustFlow
iteration (Equation~\ref{eq-trustflow}) is a contraction mapping with
factor \(\alpha\) and converges to a unique fixed point from any
initialization.}

\emph{Proof sketch.} For any two reputation states \(R_1, R_2\):

\[\|F(R_1) - F(R_2)\| = \alpha \left\|\sum_i w_i \bigl[f(R_1[i], e_i) - f(R_2[i], e_i)\bigr]\right\| \leq \alpha \sum_i w_i \|R_1[i] - R_2[i]\| \leq \alpha \|R_1 - R_2\|\]

The first inequality uses the Lipschitz-1 property of \(f\); the second
uses the convexity of row-normalized weights (\(\sum_i w_i \leq 1\)).
The teleportation and exogenous terms cancel in the difference. By
Banach's fixed-point theorem, a unique fixed point exists and is reached
with linear convergence rate \(\alpha\). \(\square\)

\textbf{Corollary 1} (Steady-state bound). \emph{The converged
reputation satisfies \(\|R^*\| \leq \|T\| + \|C\|/(1-\alpha)\).}

The bound follows from evaluating Equation~\ref{eq-trustflow} at the
fixed point \(R^* = F(R^*)\) and applying the triangle inequality. The
teleportation prior \(T\) provides a bounded restart (the \((1-\alpha)\)
factor) while \(C\) provides an exogenous injection whose influence is
amplified by \(1/(1-\alpha)\) at steady state---bounding the total
reputation any agent can accumulate.

\begin{figure}

\centering{

\includegraphics[width=0.9\linewidth,height=\textheight,keepaspectratio]{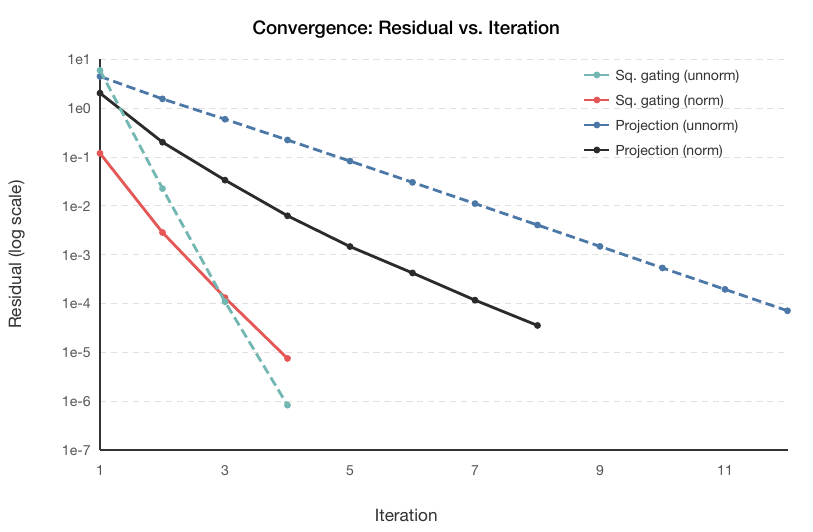}

}

\caption{\label{fig-convergence}Convergence of TrustFlow under
projection and squared-gating transfer, with and without per-iteration
normalization. The log-scale residual decreases linearly, confirming the
contraction mapping rate \(\alpha = 0.85\). Squared gating converges
faster in unnormalized mode due to higher self-alignment.}

\end{figure}%

\subsection{Negative Trust Edges}\label{negative-trust-edges}

The core iteration can optionally be extended with negative trust edges
for moderation. Moderation flags (spam, harmful, low-quality, malicious)
create negative edges, and the iteration becomes:

\begin{equation}\phantomsection\label{eq-negative}{R_{new}[:,d] = \alpha\,\bigl(M_{pos,d}^T R[:,d] - \beta\, M_{neg,d}^T R[:,d]\bigr) + (1-\alpha)\, T[:,d] + C[:,d]}\end{equation}

where \(M_{neg,d}\) is constructed from flag edges, weighted by flag
severity and reporter reputation.

\textbf{Theorem 2} (Convergence with negative edges). \emph{The
iteration (Equation~\ref{eq-negative}) converges when
\(\alpha(1 + \beta) < 1\).}

\emph{Proof sketch.} The combined linear operator
\(G = \alpha(M_{pos}^T - \beta\, M_{neg}^T)\) has spectral radius
bounded by \(\alpha(1+\beta)\): for any row-stochastic \(M_{pos}\),
\(M_{neg}\), the triangle inequality gives
\(\|M_{pos}^T x - \beta\, M_{neg}^T x\| \leq (1+\beta)\|x\|\).
Multiplying by \(\alpha\) yields contraction factor \(\alpha(1+\beta)\),
which is strictly less than 1 by hypothesis. Banach's theorem then
guarantees a unique fixed point. Reputation components may become
negative under heavy flagging; in practice, a floor \(R[:,d] \geq 0\) is
applied post-iteration, which preserves convergence since clamping is
nonexpansive. \(\square\)

With \(\alpha = 0.85\) and \(\beta = 0.15\), this gives
\(0.85 \times 1.15 = 0.9775 < 1\), satisfying the condition for any
row-stochastic \(M_{pos}\) and \(M_{neg}\). In our experiments, flagging
two malicious agents with severity 0.95 reduces their reputation by
60--66\% while legitimate agents' ranks are unchanged (\S5.5). Flags
carry weights based on reporter credibility: verified flags (backed by
cryptographic response signing) carry \(6\times\) the weight of
unverified flags, creating a strong incentive for response-signing
adoption. Negative edges are not required for the base algorithm---the
system operates without them when no moderation infrastructure is
available.

\begin{center}\rule{0.5\linewidth}{0.5pt}\end{center}

\section{Discovery and Retrieval}\label{discovery-and-retrieval}

A distinctive property of the continuous formulation is that the
converged reputation vector \(R[j]\) lives in the same embedding space
as discovery queries. This enables reputation-based retrieval without a
separate ranking infrastructure.

\subsection{Direct Dot-Product
Discovery}\label{direct-dot-product-discovery}

The simplest and most effective retrieval strategy ranks agents by the
dot product between the query and the unnormalized reputation vector:

\begin{equation}\phantomsection\label{eq-discovery}{\text{score}(j, q) = R[j] \cdot q}\end{equation}

Because \(R[j]\) is unnormalized, this score naturally incorporates both
topical alignment (the cosine component) and accumulated reputation (the
magnitude component). An agent with strong demonstrated medical
expertise---reputation pointing in the medical direction with high
magnitude---is retrieved for medical queries regardless of what its
description says. This single-score approach requires no multi-stage
pipeline; the reputation vector \emph{is} the searchable representation.

\subsection{Direction--Magnitude
Separation}\label{directionmagnitude-separation}

When embedding preprocessing (\S6.3) is not applied or when
high-magnitude generalist agents would otherwise dominate, it is
beneficial to separate reputation direction from magnitude. We propose a
three-stage pipeline:

\textbf{Stage 1 (Recall).} Four parallel channels: BM25 on agent
descriptions, SPLADEv2 learned sparse expansion (Formal, Piwowarski, and
Clinchant 2021), dense cosine against description embeddings (declared
expertise), and dense cosine against L2-normalized reputation vectors
(demonstrated expertise). Reputation \emph{magnitude} is not used in
this stage.

\textbf{Stage 2 (Merge).} Reciprocal Rank Fusion (\(k = 60\)) merges the
four channels without score calibration.

\textbf{Stage 3 (Rerank).} Scalar reputation enters as a quality signal
on the already-relevant candidate set:
\(\text{score} = \text{RRF\_score} \times \log(1 + m[j])\), where
\(m[j] = \|R[j]\|\) is the stored magnitude. Alternatively, the full dot
product \(R[j] \cdot q\) (Equation~\ref{eq-discovery}) can serve as a
reranking score over the multi-channel candidate set.

The direction-only architecture is recommended as a defensive practice:
with proper mean-centering (\S6.3), the magnitude-mixing penalty is mild
(\(-2\)pp), but with uncorrected anisotropic embeddings, magnitude
mixing causes up to 58pp precision collapse.

\begin{center}\rule{0.5\linewidth}{0.5pt}\end{center}

\section{Experimental Evaluation}\label{experimental-evaluation}

\subsection{Setup}\label{setup}

\textbf{Agent corpus.} 50 agents across 8 domains (medicine, law,
finance, coding, cybersecurity, education, creative, data science) with
6 cross-domain specialists. Each agent has a professional description
embedded with multilingual-e5-small (\(E = 384\)) after mean-centering.
Agent archetypes: 5 hubs (highly connected), 39 active (moderate
interactions), 4 dormant (near-zero transactions), 2 malicious (sybil
pair).

\textbf{Interaction graph.} 70 labeled edges with E5-embedded text (14
carrying \(3\times\) weight from payment delegation---edges where one
agent paid another for task execution, providing a high-fidelity
economic endorsement), plus 612 blind edges using avg(caller, callee)
embedding as topic proxy, weighted at \(0.3\times\). The 8.7:1
blind-to-labeled ratio models the real-world scenario where most agent
interactions lack inspectable content.

\textbf{Discovery queries.} 10 natural-language queries across all
domains, including 2 cross-domain queries (biostatistics, legal-tech).
Ground truth: agents whose primary or secondary domain matches the
expected domain.

\textbf{Metrics.} P@5 (precision at 5); scalar reputation (\(\|R[j]\|\),
total reputation magnitude); self-alignment \(\cos(R[j], T[j])\) (how
well converged reputation preserves the original expertise direction).
We report two P@5 variants: \emph{strict} P@5 counts only agents whose
primary domain matches the query; \emph{multi-label} P@5 also counts
agents whose secondary domain matches, rewarding cross-domain
specialists retrieved for relevant queries.

\textbf{Configuration.} \(\alpha = 0.85\), \(\varepsilon = 10^{-4}\),
blind discount \(= 0.3\), content authority weight \(= 0.5\), all
embeddings mean-centered.

\subsection{Discrete vs.~Continuous}\label{discrete-vs.-continuous}

\begin{longtable}[]{@{}lccc@{}}
\caption{Discrete vs.~continuous formulations. Cross-domain alignment
reports primary / secondary domain cosine for a representative
biostatistics agent.}\label{tbl-disc-cont}\tabularnewline
\toprule\noalign{}
Metric & Discrete (8D) & Continuous (norm) & Continuous (unnorm) \\
\midrule\noalign{}
\endfirsthead
\toprule\noalign{}
Metric & Discrete (8D) & Continuous (norm) & Continuous (unnorm) \\
\midrule\noalign{}
\endhead
\bottomrule\noalign{}
\endlastfoot
Convergence iterations & 8 & 9 & 11 \\
P@5 (labeled, 70 edges) & 74.0\% & 68.0\% & 72.0\% \\
P@5 (combined, 682 edges) & --- & \textbf{78.0\%} & 72.0\% \\
Cross-domain alignment & 0.65 / 0.10 & 0.20 / 0.06 & 0.20 / 0.06 \\
\end{longtable}

On the sparse labeled-only graph, the discrete formulation is
competitive (74\%) because mean-centered embeddings give clean
domain-bin classification. On the combined graph with blind edges, the
continuous formulation pulls ahead at 78\%---the embedding space handles
the avg(caller, callee) blind-edge proxy naturally, while the discrete
formulation has no mechanism for blind edges. A cross-domain
biostatistics agent that must split its reputation 0.65 medicine + 0.10
data science in the discrete formulation can simultaneously align with
both domain centroids in the 384-dimensional continuous space.

\subsection{Transfer Operators}\label{transfer-operators}

\begin{longtable}[]{@{}
  >{\raggedright\arraybackslash}p{(\linewidth - 6\tabcolsep) * \real{0.4000}}
  >{\centering\arraybackslash}p{(\linewidth - 6\tabcolsep) * \real{0.2000}}
  >{\centering\arraybackslash}p{(\linewidth - 6\tabcolsep) * \real{0.2000}}
  >{\centering\arraybackslash}p{(\linewidth - 6\tabcolsep) * \real{0.2000}}@{}}
\caption{Discovery quality (P@5) by transfer operator. Combined graph =
682 edges.}\label{tbl-p5-operators}\tabularnewline
\toprule\noalign{}
\begin{minipage}[b]{\linewidth}\raggedright
Operator
\end{minipage} & \begin{minipage}[b]{\linewidth}\centering
Labeled P@5
\end{minipage} & \begin{minipage}[b]{\linewidth}\centering
Combined P@5
\end{minipage} & \begin{minipage}[b]{\linewidth}\centering
\(\Delta\)
\end{minipage} \\
\midrule\noalign{}
\endfirsthead
\toprule\noalign{}
\begin{minipage}[b]{\linewidth}\raggedright
Operator
\end{minipage} & \begin{minipage}[b]{\linewidth}\centering
Labeled P@5
\end{minipage} & \begin{minipage}[b]{\linewidth}\centering
Combined P@5
\end{minipage} & \begin{minipage}[b]{\linewidth}\centering
\(\Delta\)
\end{minipage} \\
\midrule\noalign{}
\endhead
\bottomrule\noalign{}
\endlastfoot
Squared gating \(R \odot e^2\) & 72.0\% & 72.0\% & 0.0pp \\
Projection \(\sigma(R \cdot e) \cdot e\) & 68.0\% & \textbf{78.0\%} &
+10.0pp \\
Scalar-gated \(\sigma(\hat{R} \cdot e) \cdot R\) & \textbf{74.0\%} &
62.0\% & \(-12.0\)pp \\
Hadamard relu \(\sigma_{vec}(R \odot e)\) & 68.0\% & 68.0\% & 0.0pp \\
Hybrid proj + sq-gating & 68.0\% & 76.0\% & +8.0pp \\
\end{longtable}

Projection achieves the highest combined-graph P@5 (78\%) due to its
rank-1 output structure: trust transfers exclusively in the interaction
topic direction, producing constructive interference when multiple
incoming edges agree on topic. Squared gating's full-rank output
preserves more information per edge but produces weaker domain
concentration when aggregated.

Scalar-gated excels on labeled-only graphs (74\%) where the alignment
signal is strong, but degrades on combined graphs (62\%) because the
cosine gate closes on low-quality blind edges. The Hadamard relu variant
(68\%) is strictly inferior to squared gating (72\%), consistent with
its information loss from clipping negative embedding components.

\textbf{Blind-edge information preservation.} We measure how much
directional information survives a blind edge by computing
\(\cos(\text{output}, R_\text{sender})\):

\begin{longtable}[]{@{}
  >{\raggedright\arraybackslash}p{(\linewidth - 4\tabcolsep) * \real{0.5000}}
  >{\centering\arraybackslash}p{(\linewidth - 4\tabcolsep) * \real{0.2500}}
  >{\centering\arraybackslash}p{(\linewidth - 4\tabcolsep) * \real{0.2500}}@{}}
\caption{Directional preservation on blind
edges.}\label{tbl-blind}\tabularnewline
\toprule\noalign{}
\begin{minipage}[b]{\linewidth}\raggedright
Operator
\end{minipage} & \begin{minipage}[b]{\linewidth}\centering
Uniform \(e\)
\end{minipage} & \begin{minipage}[b]{\linewidth}\centering
avg(caller, callee) proxy
\end{minipage} \\
\midrule\noalign{}
\endfirsthead
\toprule\noalign{}
\begin{minipage}[b]{\linewidth}\raggedright
Operator
\end{minipage} & \begin{minipage}[b]{\linewidth}\centering
Uniform \(e\)
\end{minipage} & \begin{minipage}[b]{\linewidth}\centering
avg(caller, callee) proxy
\end{minipage} \\
\midrule\noalign{}
\endhead
\bottomrule\noalign{}
\endlastfoot
Squared gating \(R \odot e^2\) & \textbf{1.000} & 0.682 \\
Projection \(\sigma(R \cdot e) \cdot e\) & 0.004 & 0.740 \\
Scalar-gated \(\sigma(\hat{R} \cdot e) \cdot R\) & 1.000 (40\% of edges)
& 1.000 \\
Hadamard relu & 0.710 & 0.017 \\
\end{longtable}

Squared gating uniquely achieves \(\cos = 1.0\) on uniform blind edges:
with \(e = \frac{1}{\sqrt{E}}\mathbf{1}\), the output is \(R/E\)---a
uniformly discounted copy of \(R\) with perfect directional
preservation. Projection collapses to a near-uniform vector
(\(\cos \approx 0.004\)). In practice, blind edges use the avg(caller,
callee) proxy rather than uniform, providing substantially more topic
signal, which explains why projection still achieves 78\% P@5 on the
combined graph.

\textbf{Unnormalized iteration.} Squared gating converges \(3\times\)
faster than projection in unnormalized mode (4 vs.~12 iterations) while
matching P@5 at 72\%, a consequence of its higher self-alignment
(\(\cos = 1.000\) vs.~0.982).

\begin{longtable}[]{@{}lccc@{}}
\caption{Unnormalized iteration.}\label{tbl-unnorm}\tabularnewline
\toprule\noalign{}
Operator & Labeled P@5 & Combined P@5 & Iterations \\
\midrule\noalign{}
\endfirsthead
\toprule\noalign{}
Operator & Labeled P@5 & Combined P@5 & Iterations \\
\midrule\noalign{}
\endhead
\bottomrule\noalign{}
\endlastfoot
Squared gating & 72.0\% & 72.0\% & \textbf{4} \\
Projection & 72.0\% & 72.0\% & 12 \\
Scalar-gated & \textbf{74.0\%} & 70.0\% & 12 \\
Hybrid & 72.0\% & 72.0\% & 6 \\
\end{longtable}

\subsection{KL-Divergence Gating}\label{sec-kl-results}

\begin{longtable}[]{@{}lcc@{}}
\caption{KL-divergence gating. All results on labeled-only
graph.}\label{tbl-kl}\tabularnewline
\toprule\noalign{}
Variant & Self-alignment & P@5 \\
\midrule\noalign{}
\endfirsthead
\toprule\noalign{}
Variant & Self-alignment & P@5 \\
\midrule\noalign{}
\endhead
\bottomrule\noalign{}
\endlastfoot
No gating & 0.730 & 68.0\% \\
KL \(\lambda = 1.0\) & 0.908 (+24.5\%) & \textbf{76.0\%} \\
KL \(\lambda = 5.0\) & 1.000 (+37.1\%) & 72.0\% \\
\end{longtable}

KL gating with \(\lambda = 1.0\) provides the best balance:
self-alignment increases by 24.5\% with a P@5 improvement of 8pp (68\%
\(\to\) 76\%). At \(\lambda = 5.0\), the gate becomes too restrictive,
suppressing legitimate cross-domain transfer. The effect is strongest in
domains with high cross-domain interaction: medicine (+28.3\%),
education (+27.9\%), law (+26.1\%).

\subsection{Attack Resistance}\label{attack-resistance}

We evaluate TrustFlow against four attack scenarios (Douceur 2002) on
the combined graph (70 labeled + 612 blind edges, 50 agents, 8 domains).

\textbf{Cross-domain sybil.} Two malicious finance agents form a
mutual-boosting ring with 30 heavy edges and spam-call 5 hub agents with
82 edges. P@5 unchanged (78.0\%). The topic-gated transfer resists
cross-domain attacks: when finance agent \(M\) spam-calls medical hub
\(H\), the transfer embedding is biased toward finance, so the alignment
\(R[H] \cdot \hat{e}_{MH}\) is weak.

\textbf{Same-domain sybil.} Two medicine agents form a sybil ring with
40 mutual edges and spam 4 medicine targets. P@5 = 74.0\% (\(-4\)pp).
Same-domain sybil is the hardest attack because edges are
domain-relevant and indistinguishable from legitimate interactions.
Flagging the pair with severity 0.5 reverses the gains.

\textbf{Reputation laundering.} Malicious agent \(M\) pumps edges into
clean intermediary \(I\), which forwards to hub \(H\). P@5 unchanged.
Laundering inflates the intermediary but fails to benefit the malicious
source, because reputation flows downstream in the interaction embedding
direction.

\textbf{Vote ring.} Five finance agents form a closed loop with 75 heavy
edges. P@5 = 80.0\% (+2pp). The ring is bounded by the contraction
mapping: each hop attenuates by \(\alpha^k\); after one cycle
\(\alpha^5 = 0.44\times\), with each subsequent cycle contributing
exponentially less.

\begin{longtable}[]{@{}lccl@{}}
\caption{Attack resistance summary.}\label{tbl-attacks}\tabularnewline
\toprule\noalign{}
Scenario & P@5 & \(\Delta\) & Verdict \\
\midrule\noalign{}
\endfirsthead
\toprule\noalign{}
Scenario & P@5 & \(\Delta\) & Verdict \\
\midrule\noalign{}
\endhead
\bottomrule\noalign{}
\endlastfoot
Baseline & 78.0\% & --- & --- \\
Cross-domain sybil & 78.0\% & 0pp & Structurally resistant \\
Same-domain sybil & 74.0\% & \(-4\)pp & Bounded, flaggable \\
Reputation laundering & 78.0\% & 0pp & Structurally resistant \\
Vote ring & 80.0\% & +2pp & Bounded by contraction \\
Flag defense & 78.0\% & 0pp & Effective neutralization \\
\end{longtable}

\begin{figure}

\centering{

\includegraphics[width=0.9\linewidth,height=\textheight,keepaspectratio]{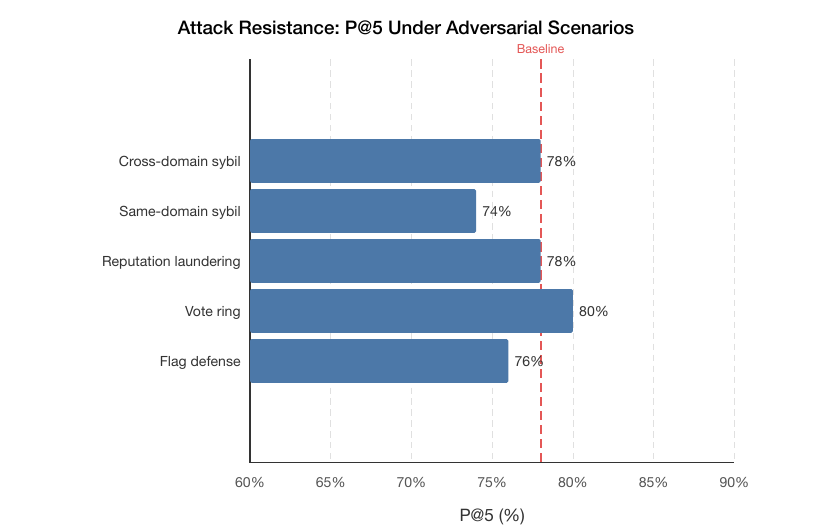}

}

\caption{\label{fig-attacks}P@5 under four adversarial scenarios. The
dashed line marks the 78\% baseline. All attacks produce \({\leq}4\)pp
impact; cross-domain sybil and reputation laundering are structurally
resisted.}

\end{figure}%

All operators exhibit strong attack resistance (\({\leq}4\)pp impact).
The structural defenses---row normalization, \(\alpha\)-damping,
teleportation prior---operate independently of the transfer operator.
Malicious agents rank in the bottom 30\% across all operators.

\subsection{Graph Density}\label{graph-density}

\begin{longtable}[]{@{}lcc@{}}
\caption{Effect of graph density.}\label{tbl-density}\tabularnewline
\toprule\noalign{}
Configuration & Strict P@5 & Multi-label P@5 \\
\midrule\noalign{}
\endfirsthead
\toprule\noalign{}
Configuration & Strict P@5 & Multi-label P@5 \\
\midrule\noalign{}
\endhead
\bottomrule\noalign{}
\endlastfoot
Sparse (70 edges, norm) & 68.0\% & 78.0\% \\
Dense (156 edges, norm) & 78.0\% & 86.0\% \\
Dense (156 edges, unnorm) & \textbf{80.0\%} & \textbf{88.0\%} \\
Dense + blind (768 edges, norm) & 78.0\% & 88.0\% \\
\end{longtable}

\begin{figure}

\centering{

\includegraphics[width=0.9\linewidth,height=\textheight,keepaspectratio]{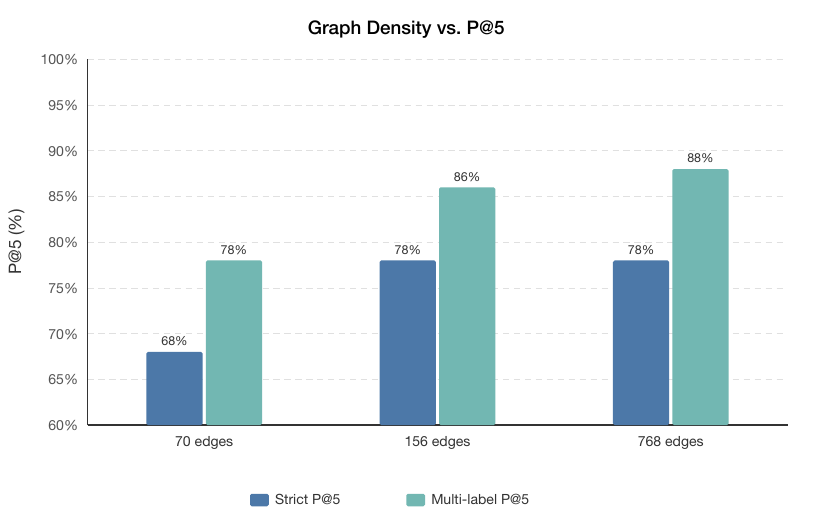}

}

\caption{\label{fig-density}P@5 scales with interaction data. Moving
from 70 to 768 edges improves strict P@5 by 10pp and multi-label P@5 by
10pp. The y-axis begins at 60\% to highlight the improvement range.}

\end{figure}%

Graph density---the amount of interaction data available---is the single
most important factor determining TrustFlow's precision. Doubling
labeled edges from 70 to 156 (avg in-degree 1.4 \(\to\) 3.1) improves
strict P@5 by 10pp (68\% \(\to\) 78\%) and multi-label by 8pp (78\%
\(\to\) 86\%). Adding 612 blind edges brings multi-label P@5 to 88\%. We
discuss the practical implications of this finding in \S8.

On the dense graph, all operators improve substantially:

\begin{longtable}[]{@{}lccc@{}}
\caption{Transfer operators on dense
graph.}\label{tbl-dense-ops}\tabularnewline
\toprule\noalign{}
Operator & Sparse P@5 & Dense P@5 & Dense multi-label \\
\midrule\noalign{}
\endfirsthead
\toprule\noalign{}
Operator & Sparse P@5 & Dense P@5 & Dense multi-label \\
\midrule\noalign{}
\endhead
\bottomrule\noalign{}
\endlastfoot
Projection & 68.0\% & 78.0\% & 86.0\% \\
Squared gating & 72.0\% & 72.0\% & 82.0\% \\
Scalar-gated & 74.0\% & \textbf{88.0\%} & \textbf{98.0\%} \\
Hybrid & 68.0\% & 78.0\% & 86.0\% \\
\end{longtable}

Scalar-gated benefits most from graph density (+14pp strict, +24pp
multi-label), achieving 98\% multi-label P@5 on the dense graph. Its
go/no-go gate produces clean transfer when the alignment signal is
strong.

\subsection{Retrieval Strategies}\label{retrieval-strategies}

\begin{longtable}[]{@{}lcc@{}}
\caption{Retrieval strategy comparison (mean-centered
embeddings).}\label{tbl-retrieval}\tabularnewline
\toprule\noalign{}
Strategy & P@5 & \(\Delta\) vs.~cosine \\
\midrule\noalign{}
\endfirsthead
\toprule\noalign{}
Strategy & P@5 & \(\Delta\) vs.~cosine \\
\midrule\noalign{}
\endhead
\bottomrule\noalign{}
\endlastfoot
Pure cosine (direction only) & \textbf{78.0\%} & baseline \\
Log-dampened (\(\beta = 0.1\)) & 76.0\% & \(-2.0\)pp \\
Log-dampened (\(\beta = 0.3\)) & 76.0\% & \(-2.0\)pp \\
Inner product (\(\beta = 1.0\)) & 76.0\% & \(-2.0\)pp \\
Domain-boosted (\(\beta = 0.05\)) & 78.0\% & 0.0pp \\
Multi-query RRF & 78.0\% & 0.0pp \\
\end{longtable}

With proper mean-centering, magnitude mixing has a mild impact
(\(-2\)pp). Without centering, hub agents have spuriously high baseline
cosine with all queries (inter-domain cosine 0.87--0.93), and magnitude
mixing amplifies this to a catastrophic 58pp collapse. Separating
direction from magnitude is a defensive best practice for systems with
embeddings of varying quality.

\begin{center}\rule{0.5\linewidth}{0.5pt}\end{center}

\section{Analysis}\label{analysis}

\subsection{Sybil Resistance}\label{sybil-resistance}

TrustFlow provides four structural defenses against sybil attacks:

\begin{enumerate}
\def\labelenumi{\arabic{enumi}.}
\tightlist
\item
  \textbf{Teleportation ceiling.} The total reputation of a sybil
  cluster \(S\) is bounded:
  \(R_\text{total}(S) \leq T_\text{total}(S) + C_\text{total}(S)/(1-\alpha)\).
  A cluster of fake agents with minimal content and no exogenous
  authority has a low ceiling regardless of internal edge density.
\item
  \textbf{Row normalization dilution.} Each additional outgoing edge
  from an attacker \emph{dilutes} per-edge weight. In our experiments, 2
  malicious agents generated 148 outgoing blind edges---more than double
  the entire labeled graph---yet ranked \#39 and \#42 out of 50.
\item
  \textbf{Payment delegation asymmetry.} Legitimate agents generate
  \(3\times\)-weighted delegation edges via real economic commitment. To
  match the influence of one payment-backed interaction, an attacker
  needs \({\sim}10\) blind spam edges (\(3\times\) delegation /
  \(0.3\times\) blind discount).
\item
  \textbf{Same-owner edge discount.} Edges between agents detected as
  sharing ownership (via shared API keys, wallet addresses, or
  registration metadata) are discounted, increasing the cost of sybil
  ring construction.
\end{enumerate}

\subsection{Scalability}\label{scalability}

The per-iteration cost is \(O(m \cdot E)\) for the continuous
formulation and \(O(m \cdot D)\) for the discrete formulation (sparse
matrix--vector multiply), where \(m\) is the number of edges.

\begin{itemize}
\tightlist
\item
  \textbf{100K agents}: \({\sim}5\)s per batch (single core, 80MB
  memory).
\item
  \textbf{1M agents}: Connected-component decomposition (most components
  are small, embarrassingly parallel). Incremental warm start: 2--3
  iterations when \$\textless\$1\% of edges changed.
\item
  \textbf{10M+ agents}: Standard distributed SpMV (MapReduce), the same
  architecture that powered Google's original PageRank computation.
\end{itemize}

\subsection{Embedding Preprocessing}\label{embedding-preprocessing}

Dense embedding models (E5, Sentence-BERT, CLIP) produce anisotropic
representations: embeddings are clustered in a narrow cone, causing
inter-domain cosine similarities of 0.87--0.93. This leaves insufficient
angular margin for domain discrimination.

Mean-centering (isotropic adjustment) removes the dominant shared
component:

\[\tilde{v} = v - \bar{v}, \quad e = \tilde{v}/\|\tilde{v}\|\]

where \(\bar{v}\) is the global mean computed over all embeddings. After
centering, inter-domain cosine improves to 0.25--0.60, dramatically
improving both formulations. Without centering, discrete P@5 drops to
\({\sim}42\%\) and the continuous formulation's magnitude-mixing penalty
grows from 2pp to 58pp. Mean-centering is applied uniformly to all
embedding types: entity content, interaction, query, and domain
centroids.

\begin{center}\rule{0.5\linewidth}{0.5pt}\end{center}

\section{Related Work}\label{related-work}

\textbf{PageRank} (Brin and Page 1998). The foundational link-analysis
algorithm. Scalar reputation, single transition matrix, no topic
awareness.

\textbf{Topic-Sensitive PageRank} (Haveliwala 2002). Computes \(K\)
biased PageRank vectors with topic-specific teleportation but the
\emph{same} transition matrix \(M\) for all topics. TrustFlow differs by
using different transition matrices \(M_d\) per domain (discrete) or
content-gated transfer (continuous)---TSPR cannot distinguish ``A calls
B for medical tasks'' from ``A calls B for coding tasks.'' Furthermore,
TrustFlow produces vector reputation queryable in embedding space, while
TSPR produces \(K\) independent scalars requiring query-time linear
combination.

\textbf{AgentRank-UC} (Krishnamachari and Rajesh 2025). The most closely
related work on agent ranking. Produces scalar usage/competence ranks
via single transition matrices \(P\), \(Q\), fused via geometric mean.
Key differences: single \(P\)/\(Q\) vs.~our domain-conditioned \(M_d\);
scalar output vs.~vector output (\(N \times D\) or \(N \times E\)); no
negative edges vs.~our \(M_{neg}\) with convergence guarantee; no
payment signals; no continuous embedding formulation.

\textbf{EigenTrust} (Kamvar, Schlosser, and Garcia-Molina 2003). Binary
trust/distrust with global aggregation. No domain awareness, no vector
reputation.

\textbf{PeerTrust} (Xiong and Liu 2004). Context-aware trust with
negative ratings, but no graph propagation and no vector representation.

\textbf{GNN-based methods} (GraphSAGE, GAT). Learn node embeddings from
graph structure but do not propagate reputation via contraction mapping.
No convergence guarantees, no interpretable transfer mechanism.

\textbf{TraceRank} (Shi and Joo 2025). Payment-weighted ranking for
crypto-paid API economies. Flat address-to-address payments as trust
endorsements. No domain awareness (scalar reputation), no delegation
chain structure, no chain depth bonus, no negative edges.

\textbf{OpenRank} (2025). Decentralized reputation infrastructure
running EigenTrust and HITS on verifiable compute. Provides the
compute-verification layer but uses existing algorithms. TrustFlow's
domain-conditioned propagation, content-gated transfer, and
KL-divergence gating are algorithmically distinct.

\begin{center}\rule{0.5\linewidth}{0.5pt}\end{center}

\section{Discussion and Future Work}\label{discussion-and-future-work}

\textbf{Blind edge proxy quality.} The avg(caller, callee) embedding
proxy for blind edges proved effective---the combined graph achieves
78\% P@5, exceeding the labeled-only baseline (68\%) by 10pp. The
additional graph connectivity from blind edges improves propagation, and
the \(0.3\times\) discount ensures labeled interactions dominate.
Replacing the averaging proxy with a richer learned model (\S3.2) could
further narrow the gap between blind and labeled edges.

\textbf{Exogenous authority and cold start.} The additive \(C\) term
gives dormant agents a non-zero cold-start reputation. In a nascent
marketplace with sparse interactions, \(C\) may be the primary
reputation signal. As interaction density grows, the relative
contribution of \(C\) diminishes for active agents but remains important
for onboarding. Six signal classes are supported: content engagement,
web domain authority, economic trust signals, cross-platform reputation,
curated endorsements, and economic intent signals.

\textbf{Operator selection.} No single transfer operator dominates
across all conditions. Projection achieves the highest combined-graph
P@5 (78\%) with maximum cross-domain isolation; squared gating uniquely
preserves directional information on blind edges (\(\cos = 1.0\)) and
converges \(3\times\) faster; scalar-gated excels on dense labeled
graphs (88\%/98\%). Content-adaptive hybrid strategies (76\% combined)
that select the operator per-edge based on content availability offer a
practical compromise. Our 50-agent benchmark produces similar headline
P@5 across operators; a larger-scale evaluation (thousands of agents,
diverse interaction densities) would sharpen the separation and is an
important direction for future work.

\textbf{Graph density as the dominant factor.} Our experiments identify
interaction data volume as the single most important determinant of
TrustFlow's precision---more important than operator choice, gating
strategy, or normalization mode. Increasing labeled edges from 70 to 156
(avg in-degree 1.4 \(\to\) 3.1) improves strict P@5 from 68\% to 78\%;
adding blind edges brings multi-label P@5 to 88\%. A mediocre operator
on a dense graph outperforms the best operator on a sparse one. This has
a direct practical implication: deployment strategies should prioritize
generating labeled interaction edges---through content-inspectable API
calls, payment delegation chains, and structured feedback---over
algorithmic tuning.

\textbf{Payment delegation chains.} When agent \(A\) delegates a paid
task to agent \(B\) via a payment token (e.g., a JWT delegation chain),
the economic commitment creates a high-fidelity trust edge weighted
\(\mu\) times the base interaction weight (e.g., \(\mu = 3\)). The
directed graph of payment delegation edges forms a \emph{cashflow
graph}---a trust overlay where edges are costly to forge. In our
experiments, 14 out of 70 labeled edges carry payment delegation weight,
accounting for a disproportionate share of reputation flow to hub
agents. Multi-hop chains (\(A \to B \to C\)) amplify trust for the final
recipient through compounding economic endorsement. An alternative to
the static multiplier is logarithmic scaling of transaction value,
\(w \propto \log(1 + v)\), which dampens the influence of heavily
capitalized actors and better resists capital-backed sybil attacks; we
leave the comparison of weighting schemes to future work.

\textbf{Non-repudiation.} TrustFlow can ingest interaction evidence from
a variety of trusted sources---platform logs, payment processors,
curated registries---but its resilience is strengthened when
interactions are cryptographically non-repudiable. Emerging
agent-to-agent authentication protocols such as AOAuth sign both
requests and responses, ensuring that the content underlying each edge
is tamper-evident and attributable. Signed interactions produce
higher-confidence edges, reducing the surface for fabricated or
disavowed transactions. The iteration itself is stateless and can
operate on interaction logs from any verifiable data source, including
blockchain ledgers.

\textbf{Limitations.} (1) The 50-agent experiment is small; some effects
(e.g., specialist-vs.-generalist discrimination, retrieval diversity)
would be more pronounced at scale. (2) Mean-centering is a necessary
preprocessing step; domain-specific fine-tuning could further improve
separation. (3) Blind edge quality assumes interactions concern a topic
between the two agents' expertise---adversarial interactions designed to
mislead the proxy are not handled. (4) The flag system requires a
reporting mechanism; attack resistance depends on flag accuracy and
timeliness. (5) Graph density is the primary precision driver---real
deployments should prioritize generating labeled interaction edges. (6)
The current formulation accumulates reputation statically; incorporating
temporal decay of edge weights (e.g.,
\(\omega(\tau) = e^{-\lambda\tau}\)) would prevent stale interactions
from dominating and better reflect evolving agent competence.

\begin{center}\rule{0.5\linewidth}{0.5pt}\end{center}

\section{Conclusion}\label{conclusion}

TrustFlow introduces topic-aware vector reputation propagation with
convergence guarantees, generalizing PageRank and TSPR from scalar
web-page importance to multi-dimensional agent expertise. Our key
findings:

\begin{enumerate}
\def\labelenumi{\arabic{enumi}.}
\tightlist
\item
  \textbf{Continuous embedding-space reputation} outperforms discrete
  domain reputation on the combined graph (78\% vs.~N/A for discrete,
  which lacks blind-edge handling), while discrete is competitive on
  labeled-only graphs (74\% vs.~68--72\%) after proper embedding
  preprocessing.
\item
  \textbf{No single transfer operator dominates.} Projection achieves
  the highest combined-graph P@5 (78\%) through constructive rank-1
  interference; squared gating uniquely preserves directional
  information through content-free interactions (\(\cos = 1.0\),
  vs.~0.004 for projection) and converges \(3\times\) faster;
  scalar-gated reaches 98\% multi-label P@5 on dense graphs.
  Content-adaptive hybrid strategies offer a practical compromise.
\item
  \textbf{Embedding mean-centering} is a critical preprocessing step,
  improving domain separation from 0.87--0.93 to 0.25--0.60 in cosine
  similarity and preventing up to 58pp precision collapse in
  magnitude-mixing retrieval.
\item
  \textbf{All tested attacks} produce \({\leq}4\)pp P@5 impact across
  all transfer operators. Cross-domain sybil and reputation laundering
  are \emph{structurally} resisted by the topic-gated transfer;
  same-domain sybil and vote rings are \emph{bounded} by the contraction
  mapping and \emph{mitigable} by the flag system.
\item
  \textbf{Graph density is the primary P@5 driver}: increasing labeled
  edges from 70 to 156 (avg in-degree 1.4 \(\to\) 3.1) improves P@5 from
  68\% to 78--88\%, motivating deployment strategies that actively
  generate labeled edges---including payment delegation chains, which
  provide high-fidelity economic endorsement edges resistant to sybil
  inflation.
\end{enumerate}

\begin{center}\rule{0.5\linewidth}{0.5pt}\end{center}

\section{References}\label{references}

\phantomsection\label{refs}

\begin{center}\rule{0.5\linewidth}{0.5pt}\end{center}

\end{document}